# Counterfactual Closing-Acceleration Risk: An Anticipatory Surrogate Safety Measure for the Blind Region of Car-Following

**Eni Solomon Laughter 1*,**

[1] School of Transportation Engineering, Chang'an University, Xi'an 710064, China;
* Correspondence: 2024134912@chd.edu.cn

**Abstract:** Surrogate safety measures allow road-safety assessment from trajectory data in which crashes are absent, yet the dominant proximity measures—time-to-collision (TTC) and its variants—assume invariant motion and are undefined whenever the following vehicle is not yet faster than its leader, so a large fraction of car-following carries no risk reading at all. This paper introduces Counterfactual Closing-Acceleration Risk (CCAR), an anticipatory surrogate measure that scores how exposed a follower is to a rear-end conflict if its leader were to brake, conditioned on the follower's current gap-closing acceleration. CCAR is evaluated on 745,540 expressway car-following frames from the SQM-W-1 trajectory dataset. Conventional measures leave 51% of frames unscored because the follower is not yet faster; across this blind region CCAR returns a graded, non-trivial risk in 98.8% of frames. CCAR is not redundant with existing measures (Spearman correlation 0.54 with modified TTC; top-decile risk-set overlap 0.19), and at fixed gap and speed its risk rises monotonically with closing acceleration. A controlled simulation with a reactive intelligent-driver-model follower and a scripted leader brake confirms that, holding gap and brake fixed, the actual collision rate rises with closing acceleration—causal support for the precursor. Parameter sensitivity, limitations, and applicability to forward-collision-warning systems are discussed.

## 1. Introduction

Road traffic crashes remain a leading cause of injury and death worldwide and rear-end collisions in particular dominate conflict counts on high-speed facilities. Because reported crashes are rare relative to the volume of traffic that produces them, and because crash records are collected only after harm has occurred, road-safety research has increasingly turned to surrogate safety measures (SSMs) which are kinematic indicators that quantify how close an interaction came to a collision without requiring a crash to have happened [1,2]. SSMs underpin proactive safety assessment by converting the abundant evidence of near-conflicts into estimates of risk, and a substantial methodological literature now formalizes how conflicts relate to crashes [3].

The practical basis for this shift has been the emergence of high-resolution vehicle trajectory data. Aerial and drone-based extraction now yields position, speed, and acceleration for every vehicle in a scene at high frequency, enabling microscopic study of interactions that loop-detector or probe data cannot resolve [4]. Such data are not without difficulty: raw extracted positions carry noise that is amplified by differentiation, so trajectory reconstruction and smoothing are prerequisites for any acceleration-based analysis [5,6]. Once reconstructed, trajectory datasets support detailed examination of the longitudinal and lateral behaviour of a vehicle that generates conflicts.

Among these behaviors, car-following governs longitudinal dynamics and is the proximate setting of rear-end risk. Decades of car-following modelling, from stimulus-response and safe-distance formulations to data-driven and probabilistic approaches, describe how a follower adjusts its motion to its leader [7,8], and a recurring theme is the heterogeneity of driver behaviour where different drivers maintain different gaps and accelerate and brake differently in otherwise identical situations [9]. A strand of this literature has long recognized that acceleration, not only spacing and relative speed, shapes following dynamics, motivating models that incorporate the leader's or follower's acceleration explicitly [10,11]. The follower's acceleration is therefore an established, measurable, and behaviorally meaningful descriptor of a following state.

Proactive safety assessment applies SSMs to these following states. Time-to-collision (TTC) and its relatives, the deceleration rate to avoid a crash (DRAC) and modified TTC (MTTC), remains the most widely used longitudinal measures, complemented by time-integrated and severity-oriented variants and by risk-surrogate and potential-field formulations [12–14]. Empirical comparisons indicate that drivers themselves rely on braking- and headway-based cues at least as much as on raw time-to-collision [15], underscoring that a useful SSM must reflect how risk actually accrues during following rather than only how imminent a collision is under present motion.

This is precisely where the dominant measures fall short. TTC presumes that both vehicles hold their instantaneous velocities, so it is defined only when the follower is already faster than its leader; when the follower is slower or matched, TTC is infinite and the state is deemed safe even if the follower is accelerating hard into the gap [16,17]. Field-theoretic measures relax this by producing continuous risk surfaces [18,19], and reaction-time and deceleration-based measures introduce the danger of abrupt braking [20], but none reads risk from the specific precursor that precedes many rear-end conflicts: a follower accelerating to absorb a comfortable gap, a maneuver documented in merging and weaving behaviour [21] and shown to contribute to conflicts during the anticipation phase of nearby maneuvers [22,23]. An anticipatory measure that reads this precursor is therefore needed.

This paper proposes Counterfactual Closing-Acceleration Risk (CCAR). The aim of the measure is to assign every car-following instant a scalar describing how exposed the follower is to a rear-end conflict *"if"* the leader were to brake now, given how hard the follower is currently closing the gap. The intuition is that the follower's gap-closing acceleration is an observable precursor of latent risk that proximity measures discards. The mechanism is a short counterfactual projection of a hypothetical leader-braking event against the follower's current motion, summarized by the minimum projected gap.

**Contributions.** This study (i) formalizes CCAR as an objective, anticipatory, counterfactual SSM for longitudinal car-following; (ii) demonstrates on real expressway data that CCAR scores a graded risk across the roughly half of following states that proximity measures cannot represent; (iii) shows that the closing-acceleration term carries risk information beyond gap and speed; and (iv) provides controlled-simulation evidence on highway-env that the closing-acceleration precursor is causally associated with collisions under leader braking. Two research questions organize the analysis: RQ1—do states exist, and how prevalent are they, in which TTC reports safety while CCAR reports risk; and RQ2—does the follower's closing acceleration carry risk information beyond what gap and speed already provide.

## 2. Problem Statement

Consider a follower at an instant where it is travelling no faster than its leader but is accelerating to close a comfortable gap. Under a constant-velocity assumption the relative speed is non-positive, so the projected gap never decreases and TTC is infinite: the state is classified safe. DRAC is likewise zero, because no deceleration is required to avoid a collision under current velocities. Yet the state is not safe in any anticipatory sense because if the leader brakes before the follower can react, the follower's accumulated closing speed and reaction lag drive the gap shut.

The problem this paper addresses is therefore the systematic blindness of proximity SSMs to risk that is latent in the follower's acceleration state. Formally, let the gap between follower and leader be g and the relative approach speed be the follower speed minus the leader speed. Proximity measures are functions of g and relative speed only; they cannot distinguish two states with identical g and identical relative speed but different follower acceleration, even though one follower is accelerating into the gap and the other is easing off **(Figure 1)**. The risk-relevant difference between these states is exactly the quantity proximity measures omit, and it is the quantity that conditional and field-based measures only partially recover [16,18,20].

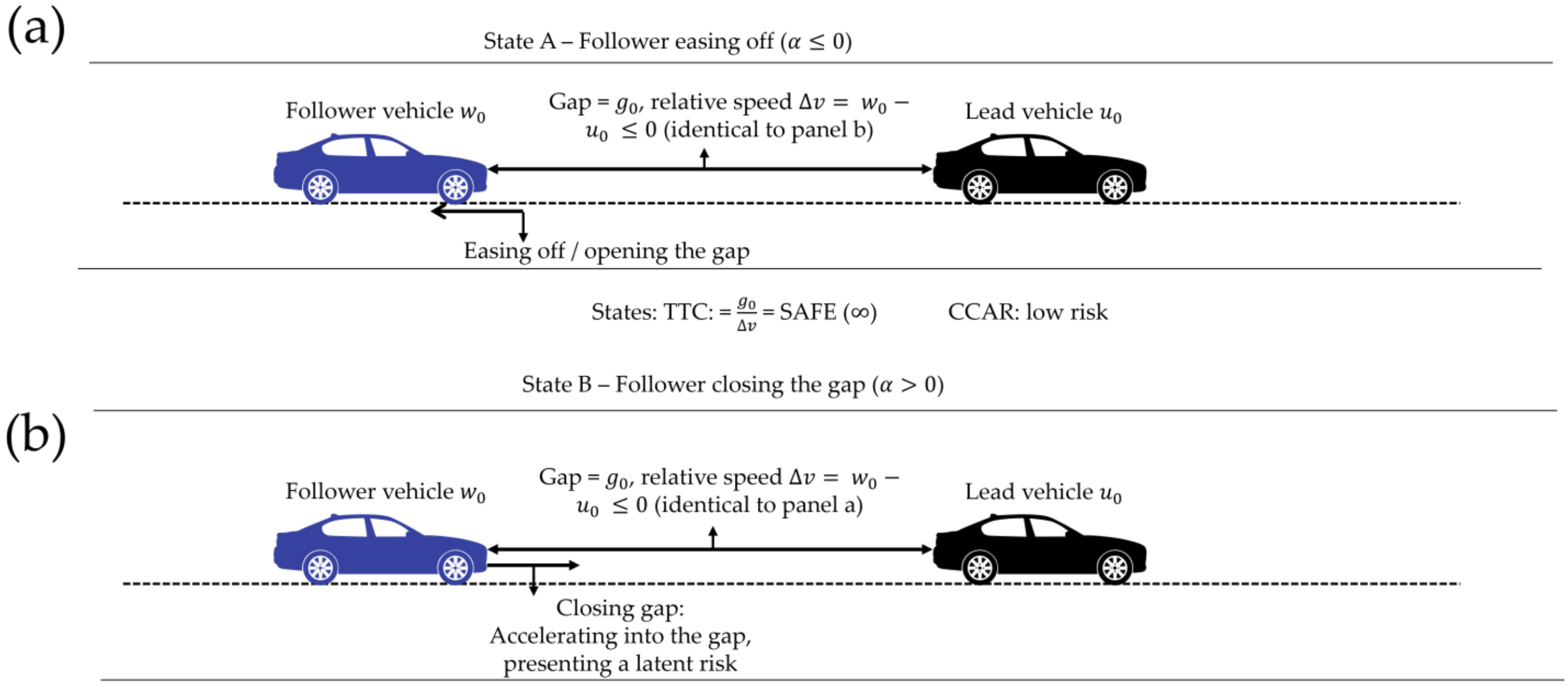

Figure 1. The blind state. Two car-following states that are indistinguishable to proximity measures (same gap, same relative speed) but carry different latent risk because the follower's acceleration differs.

A measure that resolves this blindness must satisfy three requirements: it must be defined and informative when the follower is not yet faster than the leader; it must depend on the follower's acceleration, not only on position and velocity; and it must remain interpretable as a risk score without reference to crash outcomes, because trajectory data provide none. CCAR is therefore constructed to meet these three requirements.

## 3. Related Work

*3.1. Proximity-based measures*

TTC, DRAC, and MTTC dominate longitudinal SSM practice [12,17]. They are simple and interpretable, but they require a closing relative velocity (TTC, DRAC) or an externally chosen threshold to convert a continuous quantity into a binary conflict label [17]. MTTC incorporates relative acceleration and so is defined in more states than TTC, but it remains a time-to-impact measure keyed to current relative motion rather than to a specified braking event. Extensions to two dimensions and improved conflict indicators broaden applicability to lane-change and merging interactions [24,25], yet all of these remain undefined or trivial when the follower is not yet faster than the leader—the states this paper targets.

*3.2. Field-theoretic measures*

Driving-safety-field and potential-based measures produce continuous, objective risk surfaces and can return non-zero risk where proximity measures cannot [14,19,26,27]. The conflict-field formulation of Joo et al. produces risk estimates during time windows in which other measures cannot, including in car-following [19]; the two-dimensional field indicator of Liu and Xiang explicitly criticizes the motion-state-invariance assumption that also motivates the present work [18]. These measures are the closest neighbors of CCAR. They differ in that they score a position–velocity potential rather than projecting a specific leader-braking counterfactual, and they do not single out the follower's gap-closing acceleration as the precursor signal.

*3.3. Reaction-time and deceleration-based measures*

A second family conditions risk on required or attainable deceleration and on driver reaction time. The improved risk-estimation model of Xue et al. evaluates the danger of abrupt deceleration using a safety margin that accounts for braking capability and reaction time [20]; surrogate measures of rear-end risk based on required deceleration follow similar logic [13], as do risk indices built from the heterogeneity of lane-changing behaviour [28,29]. CCAR shares the reaction-time and braking-capability ingredients but is conditioned on the follower's closing-acceleration phase in ordinary following rather than on required deceleration alone.

*3.4. Extreme value theory and the crash-free data problem*

Because trajectory data contain almost no crashes, a substantial literature uses extreme value theory (EVT) to extrapolate crash frequency from the tail of observed conflicts [3,30–32], including applications to weaving sections and intersections and to the question of how much observation is sufficient [33–35]. EVT and CCAR address the same difficulty from opposite directions: EVT extrapolates the frequency of the most severe conflicts that did occur, whereas CCAR quantifies the latent exposure of states that are not conflicts at all and therefore never enter the EVT tail. The two are complementary; CCAR is reported here alongside, instead of conventional baselines.

*3.5. Anticipation, gap-closing behaviour, and the remaining gap*

Behavioral literature documents the maneuver that motivates CCAR. Drivers approaching a merge or a slower leader frequently accelerate to claim a gap and then decelerate to synchronize [21]; the new follower's behaviour during the anticipation phase of a lane change creates and closes gaps [23,36]; the lane-changer's acceleration contributes to conflicts [22]; and asymmetric acceleration–deceleration and surrounding-traffic effects shape following risk [37–39]. After accounting for proximity, field-theoretic, reaction-time, and EVT approaches, the distinguishing element of CCAR is its counterfactual framing—scoring risk by projecting a specified leader-braking event against the follower's present closing-acceleration state, in longitudinal car-following. A deceleration-based index of this spirit (the potential index for collision with urgent deceleration) has been noted as promising but treated through correlation and simulation rather than as a developed anticipatory counterfactual [40].

## 4. The CCAR Measure

*4.1. Aim and intuition*

CCAR assigns a car-following state a scalar risk by answering a counterfactual question: if the leader were to brake now, how close would the two vehicles come, given that the follower continues its current motion through a reaction window before braking? The intuition is that a follower with positive closing acceleration is accumulating risk that is invisible to a constant-velocity reading measure; CCAR makes that accumulation explicit by projecting it forward against a hypothetical leader brake.

*4.2. Counterfactual construction*

Let the evaluation instant define a projection clock starting at zero. The measured quantities at that instant are the leader speed $u_0$, follower speed $w_0$, bumper-to-bumper gap $g_0$, and closing acceleration α (the measured follower acceleration, set to zero when negative so the measure scores only genuine closing). With three assumed parameters to complete the counterfactual model: the leader braking rate $b_L$, the follower reaction time $t_r$, and the follower braking capability $b_F$.

The leader brakes immediately at constant $b_L$ and stops at $T_L = u_0/b_L$. Its speed and displacement accounted for are:

$$u(\tau) = max(u_0 - b_L \cdot \tau, 0) \quad (1)$$

$$\Delta_L(\tau) = \begin{cases} u_0\tau - \frac{1}{2}b_L\tau^2, & \tau \le T_L, \\ \frac{u_0^2}{2b_L}, & \tau > T_L \end{cases} \quad (2)$$

The follower keeps its current acceleration through the reaction window and then brakes. With $w_1 = w_0 + \alpha t_r$ and $d_1 = w_0 t_r + \frac{1}{2}\alpha_r^2$:

Then the follower velocity and displacement;

$$w(\tau) = \begin{cases} w_0 + \alpha\tau, & \tau \le t_{r,} \\ \max(w_1 - b_F(\tau - t_r), 0), & \tau > t_r \end{cases} \quad (3)$$

$$\Delta_F(\tau) = \begin{cases} w_0\tau + \frac{1}{2}\alpha\tau^2, & \tau \le t_r \\ d_1 + w_1(\tau - t_r) - \frac{1}{2}b_F(\tau - t_r)^2, & \tau > t_r \end{cases} \quad (4)$$

The projected gap is the initial gap plus the leader's displacement minus the followers:

$$g(\tau) = g_0 + \Delta_L(\tau) - \Delta_F(\tau) \quad (5)$$

Equation (5) decomposes back into the three features the measure is built from: setting $\tau = 0$ returns $g_0$; the $+\Delta_L$ term is the leader; the $-\Delta_F$ term is the follower. The dynamics that distinguish CCAR from TTC lie in the second derivative of the gap during the reaction phase, which equals $-(b_L + \alpha)$: leader braking and follower closing acceleration enter additively, and this additive term is exactly what a constant-velocity measure discard.

*4.3. Risk read-outs and the discriminator*

The core quantity is the minimum projected gap and its normalized risk score:

$$s_{min} = \min_{\tau \ge 0} g(\tau)$$

$$R = max\left(0,1 - \frac{s_{min}}{g_0}\right) \in [0,1] \quad (6)$$

A counterfactual collision occurs when $s_{min} \le 0$; severity is read as the closing speed at the first instant the gap reaches zero. Because $g(\tau)$ is piecewise quadratic, $s_{min}$ is obtained exactly by evaluating g at a small set of candidate times—the stationary point of each moving phase and the phase boundaries—with no time discretisation. The quantity that justifies the measure is the discriminator set: states in which TTC is undefined or above a safe threshold yet CCAR reports risk. If the follower's closing acceleration carried no information beyond gap and speed, this set would be empty or would coincide with what MTTC already flags.

*4.4. Parameters as assumptions*

The three parameters $b_L$, $t_r$, and $b_F$ are modelling assumptions, not measured constants, and are treated as swept sensitivity parameters. $b_L$ is the assumed emergency braking rate of the leader; $t_r$ is the follower's perception-reaction time, the window during which it continues its current motion before braking; and $b_F$ is the follower's attainable braking capability. Each is treated as a swept sensitivity parameter rather than a fixed value. The reaction time is swept over 0.5, 1.0, and 1.5 s, values used in the trajectory-based reaction-time literature [28]; the leader and follower braking rates are swept across a comfortable-to-emergency range, with a deceleration of 5 m/s² used elsewhere as a hazardous-braking marker giving a scale reference [24]. The full factorial combination of three values for each of the three parameters produces 27 settings. No conclusion is reported that depends on a single chosen value of these parameters.

## 5. Data and Methods

*5.1. Dataset*

The analysis uses SQM-W-1 Nanjing China Site, a high-resolution aerial trajectory dataset from the Ubiquitous Traffic Eye family covering a dual-carriageway expressway segment with a constructed central barrier separating two opposing traffic streams and partial lanes that create a continuous merge-influence zone **(Figure 2)**. The dataset contains 822,712 vehicle-frame records for 1,041 vehicles across ten lanes, sampled at 24 Hz; each record provides vehicle identity, lane, longitudinal and lateral position, speed, acceleration, and leader and follower identities with gap distances. The Ubiquitous Traffic Eye family is established in the SSM and merging literature, including a macroscopic safety indicator validated on it and car-following and merging studies [41–43].

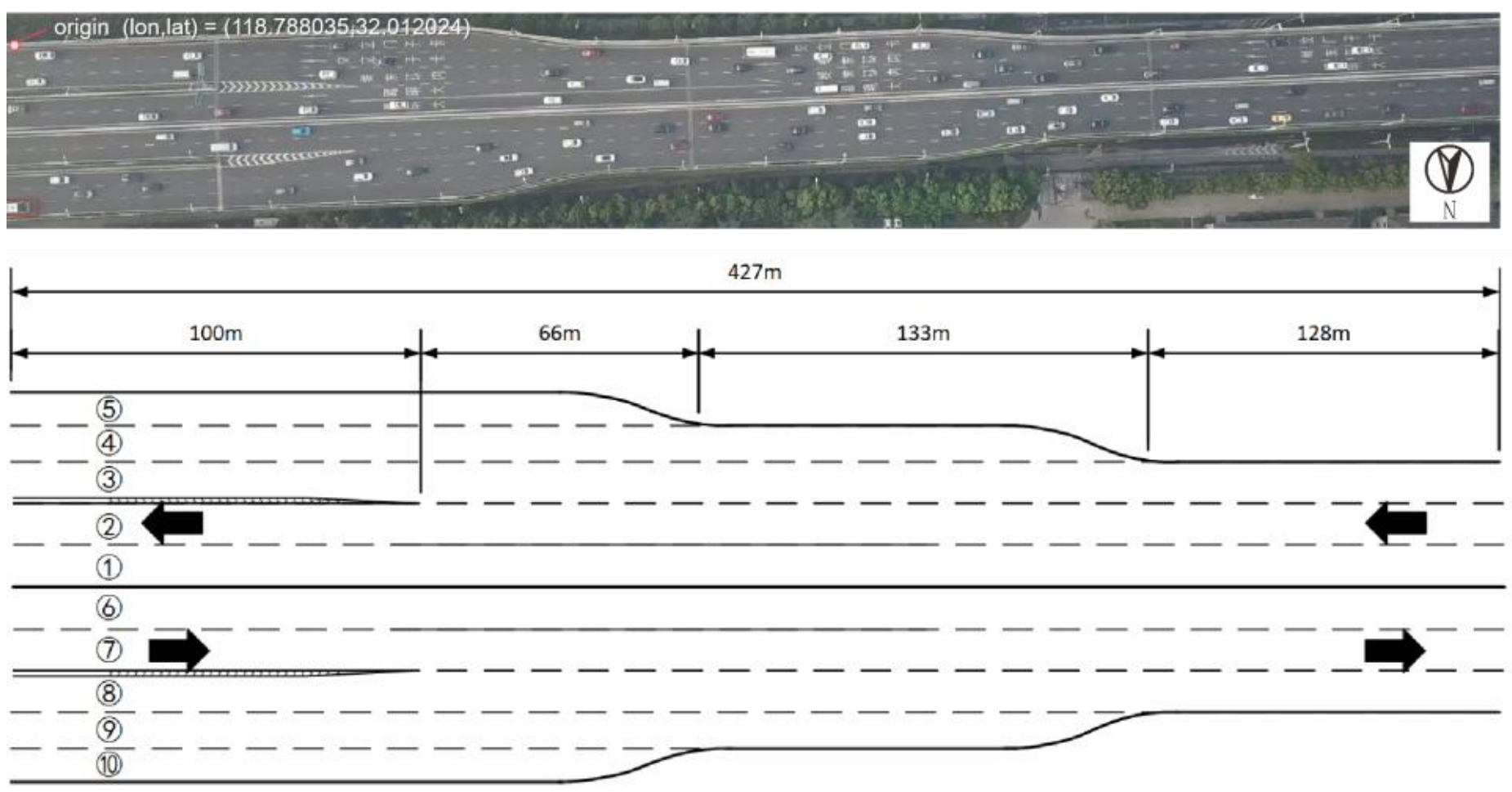

Figure 2. The SQM-W-1 study site: a dual-carriageway expressway weaving section with partial lanes forming a continuous merge-influence zone.

*5.2. Data-quality audit*

Because CCAR is acceleration-driven, the acceleration field must be trustworthy. A derivative-consistency audit compared each stored kinematic column against the numerical derivative of the column below it. The sampling interval is regular at 0.0417 s (24 Hz).

Position-derived speed matched the stored speed to a root-mean-square error RMSE of 0.052 m/s, 0.4% of the median speed of 11.96 m/s; speed-derived acceleration matched the stored acceleration to 0.134 m/s² against a median absolute acceleration of 0.493 m/s², with the stored acceleration smoother than the derived value, indicating prior filtering consistent with standard trajectory-reconstruction practice [5,6]. Implausible accelerations beyond ±8 m/s² accounted for 0.046% of frames. No dedicated reconstruction stage was required; a light guard removed the sub-percent implausible tail and non-positive gaps at the point of use.

### *5.3. Leader–follower pairing and baselines*

For this model to work a car following situation must be extracted from the dataset. Followers were paired with their leaders at each frame using the leader-identity field as provided by the dataset; frames with no leader, a non-positive gap, or an implausible acceleration were excluded, leaving 745,540 car-following frames (90.6% of the raw records). TTC, DRAC, and MTTC were computed per frame as baselines, MTTC using the leader acceleration obtained by joining each follower frame to its leader's record. Baseline thresholds were taken from the literature: TTC critical values of 1.25 s in weaving sections [33,44], 1.5 s in real-time conflict prediction [45], and 4 s for time-integrated severity [46].

### *5.4. Parameter sweep*

CCAR was computed at a central setting ($b_L$= 6 m/s², $t_r$ = 1.0 s, $b_F$ = 6 m/s²) and across a full grid of three values for each parameter (27 settings), so that every reported pattern could be tested for robustness. Correlations against MTTC were computed on closing frames where MTTC is finite.

## 6. Results

### *6.1. Existence and prevalence*

The discriminator set is non-empty and substantial. At the central setting, 14.7% of all following frames yield a counterfactual collision, and the mean normalized risk is 0.600. Among frames that TTC classifies as safe, the proportion that nonetheless produces a counterfactual collision is 14.4% at a 1.25 s threshold, 14.3% at 1.5 s, and 13.6% at 4 s—states that every proximity reading would dismiss but that close to contact if the leader brakes. The severity of these counterfactual contacts has a median closing speed of 5.4 m/s and a 90th percentile of 8.3 m/s **(Figure 3)**.

The result appears to be most striking in a state where the follower is not yet faster than the leader. Here TTC is infinite and DRAC is zero by construction, and this region comprises 51.2% of all following frames. Across it CCAR returns a graded risk that is non-zero in 98.8% of frames, with a median of 0.485 and a 90th percentile of 0.922; 7.6% of blind-region frames produce a counterfactual collision. CCAR thus provides a risk reading over the majority of car-following that proximity measures leave unscored.

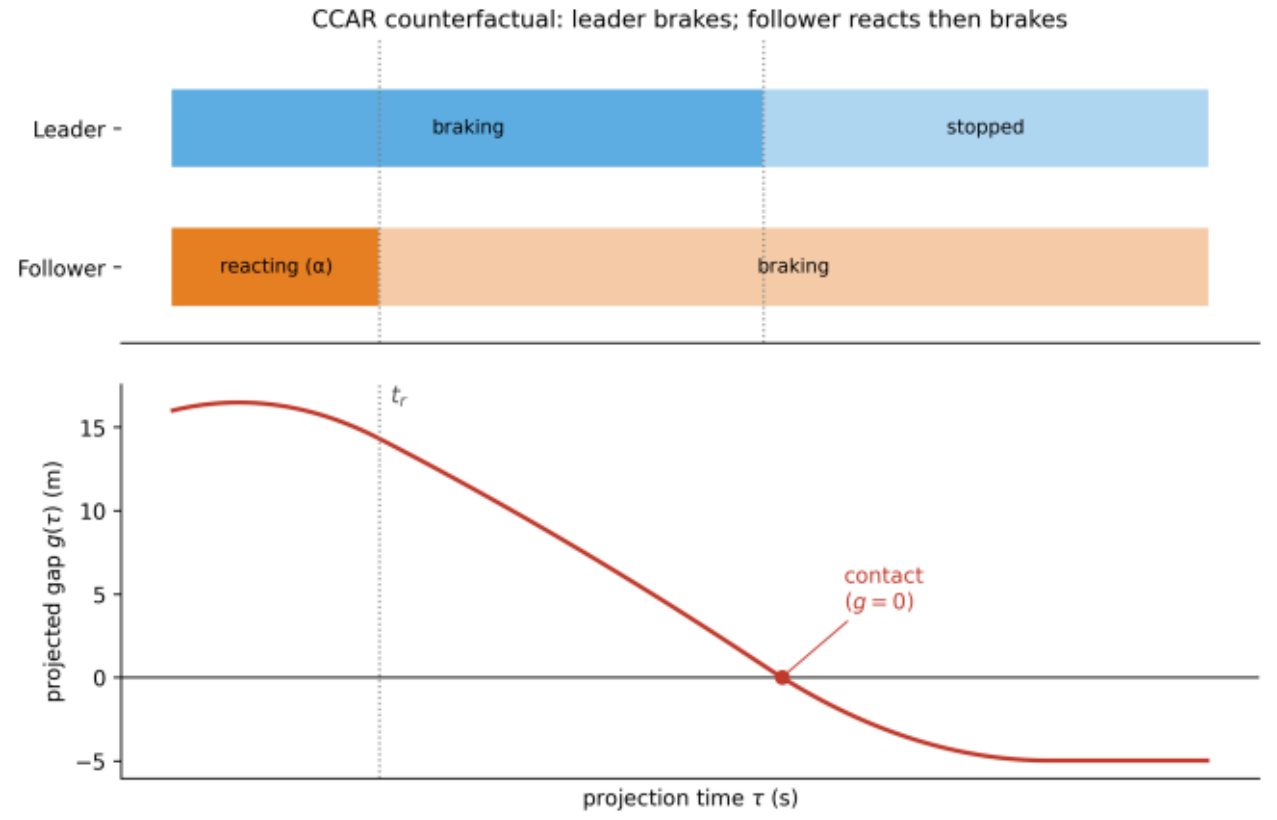

Figure 3. The CCAR counterfactual. The leader brakes while the follower continues its closing acceleration through the reaction window before braking; the projected gap $g(\tau)$ reaches a minimum $s_{min}$ marking the point of maximum exposure.

*6.2. Incremental information*

This section tests whether CCAR's risk measure carries information beyond the conventional baselines — specifically, whether the follower's closing acceleration contributes risk signal that gap and speed alone cannot explain. CCAR is not a monotone re-expression of the baselines. Across closing frames, the Spearman rank correlation between CCAR risk and the conventional measures is 0.39 for TTC, 0.25 for DRAC, and 0.54 for MTTC—correlated, as expected of risk measures, but far from redundant. The top-decile risk sets of CCAR and MTTC overlap with a Jaccard index of only 0.21, meaning that roughly four-fifths of the frames CCAR ranks most dangerous are not those MTTC ranks most dangerous.

The decisive test holds gap and speed approximately fixed and varies only by closing acceleration. Within narrow gap-by-speed cells, CCAR risk rises with closing acceleration in the substantial majority of populated cells—for example, from 0.54 to 0.91 across closing-acceleration bands at fixed gap and speed—though the rise is not strictly monotonic in every cell, and a small number of cells show a decline at the highest acceleration band. Because gap and speed are held constant, this pattern reflects information that proximity measures cannot encode, and the predominantly positive relationship answers this paper RQ2: the follower's closing acceleration carries risk information beyond gap and speed.

*6.3. Spatial characterization*

Counterfactual-collision risk is not spatially uniform. Collision rates are highest in the transition and partial lanes of one carriageway (lanes 3, 4, and 5 at 19.2%, 23.1%, and 18.9% respectively) and lower in the full-length inner lanes and the opposing carriageway (11–13%), consistent with the documented concentration of merging and lane-discovery decisions in those lanes [47] **(Figure 4)**. Collisions whose occurrence depends on the closing-acceleration term concentrate even more sharply in the same partial and transition lanes (5.1%, 3.7%, and 3.4%), matching the gap-closing maneuver that motivates the measure [21].

*6.4. Robustness across assumptions*

Across the 27 parameter settings the discriminator ranges from 0.3% to 66.4% of TTC-safe frames, with a median near 15%. A monotonic increase is observed where the discriminator rises with the assumed leader braking rate (10.3%, 22.4%, 30.3%) and reaction time (6.1%, 18.2%, 38.7%), and falls with the follower braking capability (34.5%, 17.7%, 10.9%), as physical reasoning predicts, with the extremes at the gentlest and harshest corners; it remains non-zero at every setting.

The incremental-information result holds: the Spearman correlation against MTTC stays within 0.36–0.62 across all settings, and the conditional rise of risk with closing acceleration is positive across the grid except at one extreme corner where it vanishes. The spatial ranking by mean risk is parameter-sensitive (rank correlation against the central setting has a median of 0.42), because different parameter regimes emphasize different risk mechanisms; the concentration of closing-acceleration-driven collisions is markedly more stable (median rank correlation 0.71) and is the spatial result treated as reliable.

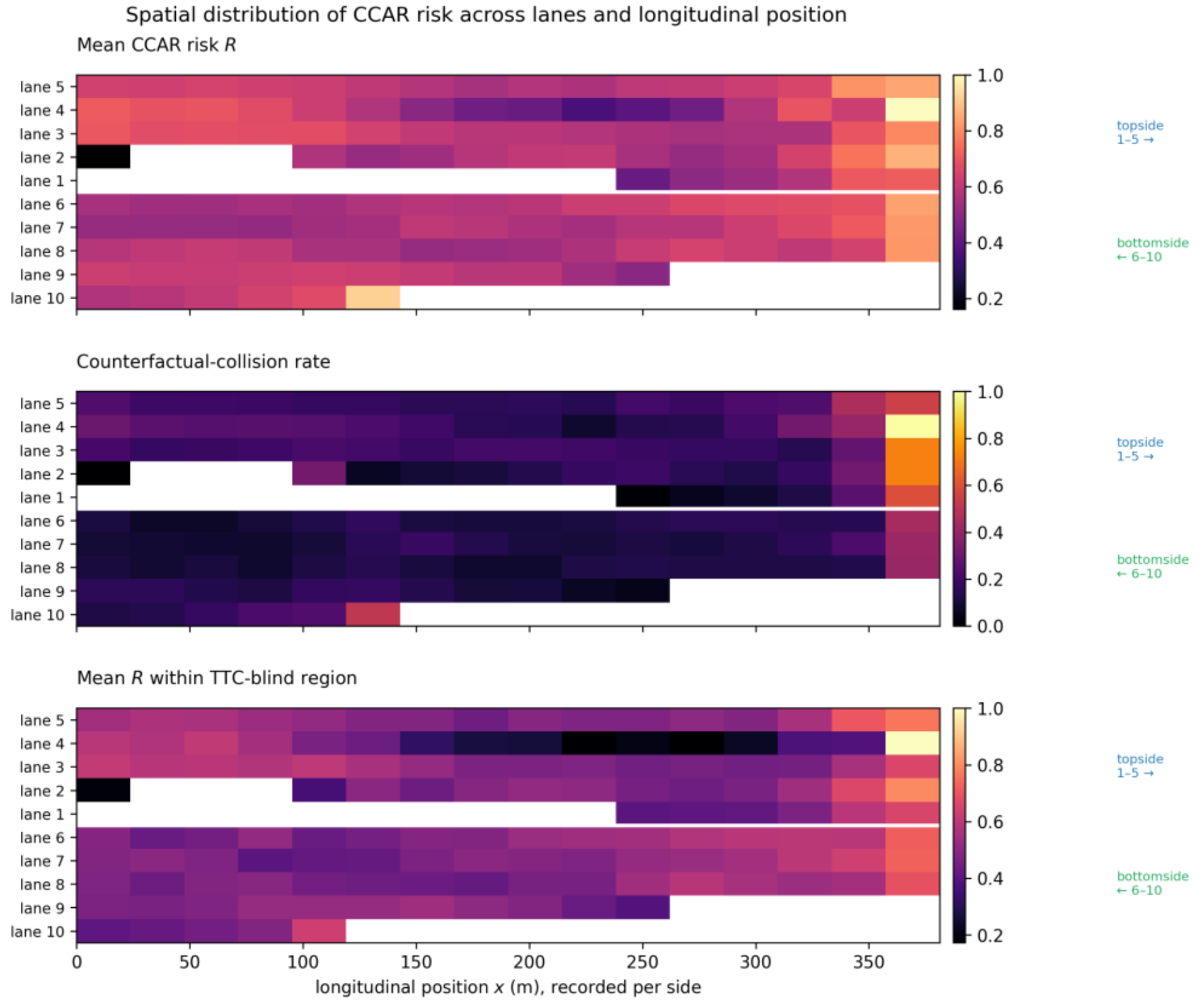

Figure 4. Spatial distribution of CCAR risks across lanes and longitudinal position; risk concentrates in the partial and transition lanes of the merge-influence zone.

## 7. Simulation Validation

### 7.1. Design

Observational trajectory data cannot manipulate the leader's braking or observe outcomes, so a controlled simulation tested whether the closing-acceleration precursor is causally associated with collisions. A single-lane two-vehicle scene was built in the highway-env simulator [48], with a follower governed by the intelligent driver model (IDM) [49] and a leader whose braking was scripted. At an evaluation instant the simulated world was duplicated; in the copy the leader executed a brake at rate $b_L$ while the IDM follower reacted naturally, and the resulting minimum gap and contact were recorded, so CCAR's prediction could be compared with the realised outcome of a genuinely reactive follower.

Driver heterogeneity was sampled across timid-to-aggressive IDM parameters so that the simulated states span the realistic range of following, including the short gaps observed in the data [9,50,51]; 700 scenarios produced 4,200 leader-brake forks. A pilot configuration using the default IDM produced no contacts at all, because the default model maintains very large, desired gaps—an instructive demonstration that car-following models with built-in collision avoidance mask risk and dominate simulation-based safety evaluation [52,53], and the reason the analysis relies on realistic heterogeneity and reports continuous minimum gaps rather than crash counts alone.

### 7.2. Causal effect of closing acceleration

Holding the gap band and the leader braking rate fixed, the actual collision rate of the reactive IDM follower rises with closing acceleration. At a leader braking rate of 6 m/s² and gaps of 12–20 m, the contact rate increases from 0.22 to 0.60 across closing-acceleration bands; the same monotone rise appears at 20–40 m gaps. Because gap and braking are fixed, this is controlled evidence that the closing-acceleration precursor is causally associated with collisions under leader braking, complementing the correlational result from the observational data **(Figure 5)**.

Figure 5. Controlled simulation: the actual collision rate of a reactive IDM follower rises with closing acceleration at fixed gap and leader braking.

### *7.3. Calibration against a reactive follower*

CCAR's predicted minimum gap tracks the realized minimum gap with a Spearman correlation of 0.59, so the ranking of danger transfers to a reactive follower. The level, however, is optimistic at the central follower braking rate: CCAR predicts contact in 16.5% of forks against an actual 30.2%, because the IDM follower brakes more gently than the assumed $b_F = 6$ m/s$^2$ and therefore closes more than the measure anticipates. This is a calibration finding, assuming a gentler follower braking rate raises CCAR's predicted contact toward the realized rate.

## 8. Discussion

### *8.1. Interpretation*

CCAR is an objective, anticipatory risk score that is informative precisely where proximity measures are silent—the majority of car-following in which the follower is not yet faster than the leader—and the follower's closing acceleration carries risk information beyond gap and speed, both correlational in the data and causally in simulation. CCAR scores the exposure of a state, not the probability that a crash will occur, and the absence of crashes in trajectory data means it cannot be validated against outcomes there.

### *8.2. Applicability to advanced driver-assistance systems*

CCAR is well suited to the ego-follower setting of forward collision warning. In that setting the host vehicle knows its own speed and acceleration directly and senses only the leader's gap and speed, so the precursor CCAR depends on—the follower's closing acceleration—is the quantity the system measures most reliably, while the sensed leader state is what mature adaptive-cruise and automatic-emergency-braking radar already provides. Behaviour-based collision-warning systems built from comparable kinematic cues are established [54]. CCAR is therefore more directly deployable in this configuration than in third-person research data. An in-vehicle study addressing sensing latency, leader-state noise, and real-time computation would be required before any operational claim.

### *8.3. Limitations*

Several limitations bound the conclusions. The study is conditional rather than predictive. The parameters $b_L$, $t_r$, and $b_F$ are assumptions; only findings that survive the parameter sweep are reported, and any single CCAR value must travel with its assumptions. The novelty is narrow: field-theoretic and reaction-time measures already occupy adjacent ground [18–20], and the distinguishing element is the closing-acceleration-conditioned counterfactual. The spatial ranking by mean risk is parameter-sensitive and is not reported as a parameter-invariant law; only the closing-acceleration-driven concentration is treated as reliable. The simulation results are conditional on the sampled driver-parameter distribution and are not yet calibrated to the dataset's own distributions. Finally, the dataset's absence of crashes provides no ground truth for this model and future works should focus on testing this model on crash related datasets.

## 9. Conclusions

This paper introduced CCAR, a counterfactual closing-acceleration risk measure for longitudinal car-following and evaluated on 745,540 expressway trajectory frames with controlled-simulation validation. CCAR scores a graded risk across the 51% of following states that proximity measures leave undefined, is not redundant with TTC, DRAC, or MTTC, and responds to the follower's closing acceleration at fixed gap and speed; controlled simulation confirms that this precursor is causally associated with collisions under leader braking. **Table 1** summarises how CCAR relates to conventional measures.

Table 1. Comparison of CCAR with conventional longitudinal surrogate safety measures.

| Property | TTC | DRAC | MTTC | CCAR |
|---|---|---|---|---|
| Underlying quantity | Time to impact | Required deceleration | Time to impact (accel-aware) | Counterfactual gap exposure |
| Defined when follower not yet faster (≈51% of frames) | No | No | Partly | Yes |
| Uses follower acceleration | No | No | Relative acceleration | Closing acceleration (precursor) |
| Anticipatory leader-brake counterfactual | No | No | No | Yes |
| Severity output | No | Yes (deceleration) | No | Yes (closing speed at contact) |
| Coverage of the blind region | None | None | Partial | Full |
| Spearman rank correlation with CCAR | 0.39 | 0.25 | 0.54 | — |
| Free of tunable parameters | No (threshold) | No (threshold) | No | No ($b_L$, $t_r$, $b_F$) |

CCAR is intended as a reusable building block: a continuous risk label that exists where conventional measures do not, a risk-aware feature for behaviour and prediction models, and a screening tool for surfacing latent-risk states. Its boundaries are equally clear as it is a longitudinal, conditional, and parameter-dependent model, and does not predict crashes or assess lateral risk. Future work will conduct more analysis on this model and release a documented implementation for public use.

Data Availability Statement: The SQM-W-1 dataset belongs to the Ubiquitous Traffic Eye family; derived analysis code supporting the reported results is available from the author on reasonable request.

Conflicts of Interest: The author declares no conflict of interest.

## References

1. Singh, D.; Das, P.; Ghosh, I. Surrogate Safety Assessment of Traffic Facilities under Ordered and Disordered Traffic Condition: Systematic Literature Review. *KSCE JOURNAL OF CIVIL ENGINEERING* **2023**, *27*, 5008–5029, doi:10.1007/s12205-023-0979-y.
2. Mustapha, A.; Abdul-Rani, A.M.; Saad, N.; Mustapha, M. Advancements in Traffic Simulation for Enhanced Road Safety: A Review. *SIMULATION MODELLING PRACTICE AND THEORY* 2024, *137*.
3. Zheng, L.; Sayed, T.; Mannering, F. Modeling Traffic Conflicts for Use in Road Safety Analysis: A Review of Analytic Methods and Future Directions. *ANALYTIC METHODS IN ACCIDENT RESEARCH* **2021**, *29*, doi:10.1016/j.amar.2020.100142.
4. Li, L.; Jiang, R.; He, Z.; Chen, X. (Michael); Zhou, X. Trajectory Data-Based Traffic Flow Studies: A Revisit. *TRANSPORTATION RESEARCH PART C-EMERGING TECHNOLOGIES* **2020**, *114*, 225–240, doi:10.1016/j.trc.2020.02.016.
5. Montanino, M.; Punzo, V. Making NGSIM Data Usable for Studies on Traffic Flow Theory Multistep Method for Vehicle Trajectory Reconstruction. *TRANSPORTATION RESEARCH RECORD* **2013**, 99–111, doi:10.3141/2390-11.
6. Ravikumar, S.; Maurya, A.; Arkatkar, S. Evaluation of Smoothing Techniques for Vehicular Trajectory Data from UAVs. In Proceedings of the Indian Institute of Technology System (IIT System); Sahu, P., Das, S., Manoj, M., Budhkar, A., Eds.; 2025; Vol. 426, pp. 351–361.
7. Yin, J.; Li, Z.; Cao, P. Systematic Review on Classical Car-Following Models. *2022 IEEE 7TH INTERNATIONAL CONFERENCE ON IN℡LIGENT TRANSPORTATION ENGINEERING, ICITE* 2022, 40–45.
8. Chen, X.; Zhang, C.; Cheng, Z.; Hou, Y.; Sun, L. A Bayesian Gaussian Mixture Model for Probabilistic Modeling of Car-Following Behaviors. *IEEE TRANSACTIONS ON IN℡LIGENT TRANSPORTATION SYSTEMS* 2024, *25*, 5880–5891.
9. Ossen, S.; Hoogendoorn, S.P. Heterogeneity in Car-Following Behavior: Theory and Empirics. *TRANSPORTATION RESEARCH PART C-EMERGING TECHNOLOGIES* 2011, *19*, 182–195.
10. Sun, B.; Wu, N.; Ge, Y.-E.; Kim, T.; Zhang, H.M. A New Car-Following Model Considering Acceleration of Lead Vehicle. *TRANSPORT* 2016, *31*, 1–10.
11. Yu, S.; Liu, Q.; Li, X. Full Velocity Difference and Acceleration Model for a Car-Following Theory. *COMMUNICATIONS IN NONLINEAR SCIENCE AND NUMERICAL SIMULATION* 2013, *18*, 1229–1234.
12. Behbahani, H.; Nadimi, N.; Naseralavi, S.S. New Time-Based Surrogate Safety Measure to Assess Crash Risk in Car-Following Scenarios. *TRANSPORTATION LETTERS-THE INTERNATIONAL JOURNAL OF TRANSPORTATION RESEARCH* 2015, *7*.
13. Zhao, P.; Lee, C. Assessing Rear-End Collision Risk of Cars and Heavy Vehicles on Freeways Using a Surrogate Safety Measure. *ACCIDENT ANALYSIS AND PREVENTION* 2018, *113*, 149–158.
14. Wu, R.; Li, L.; Shi, H.; Rui, Y.; Ngoduy, D.; Ran, B. Integrated Driving Risk Surrogate Model and Car-Following Behavior for Freeway Risk Assessment. *ACCIDENT ANALYSIS AND PREVENTION* 2024, *201*.
15. Ramezani-Khansari, E.; Moghadas Nejad, F.; Moogehi, S. Comparing Time to Collision and Time Headway as Safety Criteria. *PAMUKKALE UNIVERSITY JOURNAL OF ENGINEERING SCIENCES-PAMUKKALE UNIVERSITESI MUHENDISLIK BILIMLERI DERGISI* **2021**, *27*, 669–675, doi:10.5505/pajes.2020.79837.
16. Ge, H.; Xia, R.; Sun, H.; Yang, Y.; Huang, M. Construction and Simulation of Rear-End Conflicts Recognition Model Based on Improved TTC Algorithm. *IEEE ACCESS* 2019, *7*, 134763–134771.

17. Park, H.; Oh, C.; Moon, J.; Kim, S. Development of a Lane Change Risk Index Using Vehicle Trajectory Data. *ACCIDENT ANALYSIS AND PREVENTION* **2018**, *110*, 1–8, doi:10.1016/j.aap.2017.10.015.
18. Liu, Z.; Xiang, Q. F-F Diagram: A Two-Dimensional Surrogate Safety Indicator Based on Field Theory for Lane Change Risk Assessment. *IEEE TRANSACTIONS ON IN℡LIGENT TRANSPORTATION SYSTEMS* **2025**, *26*, 10694–10709, doi:10.1109/TITS.2025.3547969.
19. Joo, Y.-J.; Kim, E.-J.; Kim, D.-K.; Park, P.Y. A Generalized Driving Risk Assessment on High-Speed Highways Using Field Theory. *ANALYTIC METHODS IN ACCIDENT RESEARCH* **2023**, *40*, doi:10.1016/j.amar.2023.100303.
20. Xue, Q.; Wang, K.; Lu, J.J.; Xing, Y.; Gu, X.; Zhang, M. An Improved Risk Estimation Model of Lane Change Using Naturalistic Vehicle Trajectories. *JOURNAL OF TRANSPORTATION SAFETY & SECURITY* 2023, *15*, 963–986.
21. Wan, X.; Jin, P.; Yang, F.; Ran, B. Merging Preparation Behavior of Drivers: How They Choose and Approach Their Merge Positions at a Congested Weaving Area. *JOURNAL OF TRANSPORTATION ENGINEERING* **2016**, *142*, doi:10.1061/(ASCE)TE.1943-5436.0000864.
22. Chen, K.; Yu, H.; Liu, P.; Li, Z.; Wang, Y. Factors Influencing Lane-Changing Crashes and Conflicts Using Vehicle Trajectories before Crashes. *Journal of Transportation Safety & Security* **2025**, *17*, 1195–1220, doi:10.1080/19439962.2025.2509932.
23. Chen, K.; Knoop, V.L.; Liu, P.; Li, Z.; Wang, Y. Modeling the Impact of Lane-Changing's Anticipation on Car-Following Behavior. *Transportation Research Part C: Emerging Technologies* **2023**, *150*, 104110, doi:10.1016/j.trc.2023.104110.
24. Chen, K.; Li, Z.; Liu, P.; Knoop, V.L.; Han, Y.; Jiao, Y. Evaluating the Safety and Efficiency Impacts of Forced Lane Change with Negative Gaps Based on Empirical Vehicle Trajectories. *Accident Analysis & Prevention* **2024**, *203*, 107622, doi:10.1016/j.aap.2024.107622.
25. Jiang, R.; Zhu, S.; Chang, H.; Wu, J.; Ding, N.; Liu, B.; Qiu, J. Determining an Improved Traffic Conflict Indicator for Highway Safety Estimation Based on Vehicle Trajectory Data. *SUSTAINABILITY* **2021**, *13*, doi:10.3390/su13169278.
26. Chen, T.; Shi, X.; Wong, Y. A Lane-Changing Risk Profile Analysis Method Based on Time-Series Clustering. *PHYSICA A-STATISTICAL MECHANICS AND ITS APPLICATIONS* **2021**, *565*, doi:10.1016/j.physa.2020.125567.
27. Wang, B.; Chen, T.; Zhang, C.; Wong, Y.D.; Zhang, H.; Zhou, Y. Toward Safer Highway Work Zones: An Empirical Analysis of Crash Risks Using Improved Safety Potential Field and Machine Learning Techniques. *ACCIDENT ANALYSIS AND PREVENTION* 2024, *194*.
28. Chen, Q.; Gu, R.; Huang, H.; Lee, J.; Zhai, X.; Li, Y. Using Vehicular Trajectory Data to Explore Risky Factors and Unobserved Heterogeneity during Lane-Changing. *ACCIDENT ANALYSIS AND PREVENTION* **2021**, *151*, doi:10.1016/j.aap.2020.105871.
29. Chen, Q.; Huang, H.; Li, Y.; Lee, A.; Long, K.; Gu, R.; Zhai, X. Modeling Accident Risks in Different Lane-Changing Behavioral Patterns. *ANALYTIC METHODS IN ACCIDENT RESEARCH* **2021**, *30*, doi:10.1016/j.amar.2021.100159.
30. Zheng, L.; Sayed, T. A Full Bayes Approach for Traffic Conflict-Based before-after Safety Evaluation Using Extreme Value Theory. *ACCIDENT ANALYSIS AND PREVENTION* **2019**, *131*, 308–315, doi:10.1016/j.aap.2019.07.014.
31. Hussain, F.; Li, Y.; Arun, A.; Haque, M. A Hybrid Modelling Framework of Machine Learning and Extreme Value Theory for Crash Risk Estimation Using Traffic Conflicts. *ANALYTIC METHODS IN ACCIDENT RESEARCH* **2022**, *36*, doi:10.1016/j.amar.2022.100248.
32. Wang, C.; Xu, C.; Xia, J.; Qian, Z.; Lu, L. A Combined Use of Microscopic Traffic Simulation and Extreme Value Methods for Traffic Safety Evaluation. *TRANSPORTATION RESEARCH PART C-EMERGING TECHNOLOGIES* **2018**, *90*, 281–291, doi:10.1016/j.trc.2018.03.011.

33. Hu, X.; Zhou, J.; Yang, Y.; Chen, Q.; Zhang, L. Assessing the Collision Risk of Mixed Lane-Changing Traffic in the Urban Inter-Tunnel Weaving Section Using Extreme Value Theory. *ACCIDENT ANALYSIS AND PREVENTION* **2024**, *200*, doi:10.1016/j.aap.2024.107558.
34. Yue, Q.; Guo, Y.; Sayed, T.; Liu, P.; Lyu, H.; Fan, W. A Spatial Generalized Extreme Value Framework for Traffic Conflict Using Max-Stable Process Approach. *ACCIDENT ANALYSIS AND PREVENTION* **2025**, *220*, doi:10.1016/j.aap.2025.108164.
35. Paul, A.; Gore, N.; Arkatkar, S.; Joshi, G.; Haque, M. Critical Conflict Probability: A Novel Risk Measure for Quantifying Intensity of Crash Risk at Unsignalized Intersections. *IATSS RESEARCH* **2025**, *49*, 49–59, doi:10.1016/j.iatssr.2025.01.001.
36. Chen, K.; Knoop, V.L.; Liu, P.; Li, Z.; Wang, Y. How Gaps Are Created during Anticipation of Lane Changes. *Transportmetrica B: Transport Dynamics* **2023**, *11*, 958–978, doi:10.1080/21680566.2022.2152129.
37. Jokhio, S.; Duerr, M.; Bargman, J.; Baumann, M. Influence of Surrounding Traffic on Lane Change Dynamics: Insights from a Video-Based Laboratory Study. *TRANSPORTATION RESEARCH PART F-TRAFFIC PSYCHOLOGY AND BEHAVIOUR* 2024, *105*, 87–98.
38. Oh, S.; Yeo, H. Impact of Stop-and-Go Waves and Lane Changes on Discharge Rate in Recovery Flow. *TRANSPORTATION RESEARCH PART B-METHODOLOGICAL* **2015**, *77*, 88–102, doi:10.1016/j.trb.2015.03.017.
39. Xing, Y.; Wu, Y.; Wang, H.; Wang, L.; Li, L.; Peng, Y. Failed Lane-Changing Detection and Prediction Using Naturalistic Vehicle Trajectories. *TRANSPORTATION RESEARCH PART C-EMERGING TECHNOLOGIES* **2025**, *170*, doi:10.1016/j.trc.2024.104939.
40. Pinnow, J.; Masoud, M.; Elhenawy, M.; Glaser, S. A Review of Naturalistic Driving Study Surrogates and Surrogate Indicator Viability within the Context of Different Road Geometries. *ACCIDENT ANALYSIS AND PREVENTION* **2021**, *157*, doi:10.1016/j.aap.2021.106185.
41. Ye, W.; Xu, Y.; Shi, X.; Shiwakoti, N.; Ye, Z.; Zheng, Y. A Macroscopic Safety Indicator for Road Segment: Application of Entropy Theory. *PHYSICA A-STATISTICAL MECHANICS AND ITS APPLICATIONS* **2024**, *642*, doi:10.1016/j.physa.2024.129787.
42. He, Y.; Chen, J.; Yang, W.; Rao, M. Integrated Driving Behavior Modeling for Expressway Merging Zones Based on Virtual Vehicles. *PHYSICA A-STATISTICAL MECHANICS AND ITS APPLICATIONS* **2026**, *688*, doi:10.1016/j.physa.2026.131385.
43. Geng, M.; Li, J.; Xia, Y.; Chen, X. (Michael) A Physics-Informed Transformer Model for Vehicle Trajectory Prediction on Highways. *Transportation Research Part C: Emerging Technologies* **2023**, *154*, 104272, doi:10.1016/j.trc.2023.104272.
44. Zhao, Y.; Zhou, J.; Zhao, C.; Li, M. Traffic Risk Assessment of Lane-Changing Process in Urban Inter-Tunnel Weaving Segment. *TRANSPORTATION RESEARCH RECORD* **2023**, *2677*, 95–106, doi:10.1177/03611981231160171.
45. Yuan, C.; Li, Y.; Huang, H.; Wang, S.; Sun, Z.; Li, Y. Using Traffic Flow Characteristics to Predict Real-Time Conflict Risk: A Novel Method for Trajectory Data Analysis. *Analytic Methods in Accident Research* **2022**, *35*, 100217, doi:10.1016/j.amar.2022.100217.
46. Hu, Y.; Li, Y.; Huang, H. Spatio-Temporal Dynamic Change Mechanism Analysis of Traffic Conflict Risk Based on Trajectory Data. *ACCIDENT ANALYSIS AND PREVENTION* **2023**, *191*, doi:10.1016/j.aap.2023.107203.
47. Guo, Y.; Gu, X.; Chen, Y.; Guo, J.; Wan, H.; Zhou, Y. Lane Change Behavior Patterns and Risk Analysis in Expressway Weaving Areas: Unsupervised Data-Mining Method. *JOURNAL OF TRANSPORTATION ENGINEERING PART A-SYSTEMS* 2024, *150*.
48. Leurent, E. An Environment for Autonomous Driving Decision-Making 2018.
49. Treiber, M.; Hennecke, A.; Helbing, D. Congested Traffic States in Empirical Observations and Microscopic Simulations. *Phys. Rev. E* **2000**, *62*, 1805–1824, doi:10.1103/PhysRevE.62.1805.

50. Punzo, V.; Zheng, Z.; Montanino, M. About Calibration of Car-Following Dynamics of Automated and Human-Driven Vehicles: Methodology, Guidelines and Codes. *TRANSPORTATION RESEARCH PART C-EMERGING TECHNOLOGIES* **2021**, *128*, doi:10.1016/j.trc.2021.103165.
51. Punzo, V.; Montanino, M.; Ciuffo, B. Do We Really Need to Calibrate All the Parameters? Variance-Based Sensitivity Analysis to Simplify Microscopic Traffic Flow Models. *IEEE TRANSACTIONS ON INTELLIGENT TRANSPORTATION SYSTEMS* **2015**, *16*, 184–193, doi:10.1109/TITS.2014.2331453.
52. Goncu, S.; Silgu, M.A. Impacts of Car-Following Models on Simulation-Based Safety Evaluation of Freeways. *ARABIAN JOURNAL FOR SCIENCE AND ENGINEERING* 2026, *51*, 5175–5189.
53. Kendziorra, A.; Wagner, P.; Toledo, T. A Stochastic Car Following Model. *INTERNATIONAL SYMPOSIUM ON ENHANCING HIGHWAY PERFORMANCE (ISEHP), (7TH INTERNATIONAL SYMPOSIUM ON HIGHWAY CAPACITY AND QUALITY OF SERVICE, 3RD INTERNATIONAL SYMPOSIUM ON FREEWAY AND TOLLWAY OPERATIONS)* 2016, *15*, 198–207.
54. Lee, S.; Lee, S.; Kim, M. Development of a Driving Behavior-Based Collision Warning System Using a Neural Network. *INTERNATIONAL JOURNAL OF AUTOMOTIVE TECHNOLOGY* **2018**, *19*, 837–844, doi:10.1007/s12239-018-0080-6.

# APPENDIX

TABLE A1. Summary of key quantitative results.

| Metric | Value |
|---|---|
| Sampling rate | 24 Hz (0.0417 s interval) |
| Derivative-consistency RMSE — speed | 0.052 m/s (0.4% of median speed) |
| Derivative-consistency RMSE — acceleration | 0.134 m/s$^2$ |
| Implausible-acceleration frames excluded | 0.046% |
| Car-following frames analyzed | 745,540 (90.6% of raw dataset) |
| Mean CCAR risk (central parameters) | 0.600 |
| Counterfactual-collision rate (central parameters) | 14.7% |
| Discriminator: TTC-safe frames yielding a counterfactual collision (TTC ≥ 1.25 s) | 14.4% |
| Discriminator: TTC-safe frames yielding a counterfactual collision (TTC ≥ 1.5 s) | 14.3% |
| Discriminator: TTC-safe frames yielding a counterfactual collision (TTC ≥ 4.0 s) | 13.6% |
| Frames in the TTC-blind region (follower not yet faster) | 51.2% |
| CCAR risk within the blind region — median | 0.485 |
| CCAR risk within the blind region — 90th percentile | 0.922 |
| Blind-region frames with non-zero CCAR risk | 98.8% |
| Severity of counterfactual contact — median closing speed | 5.4 m/s |
| Severity of counterfactual contact — 90th percentile | 8.3 m/s |
| Spearman correlation, CCAR vs. TTC | 0.39 |
| Spearman correlation, CCAR vs. DRAC | 0.25 |
| Spearman correlation, CCAR vs. MTTC | 0.54 |
| Top-decile risk-set overlap, CCAR vs. MTTC (Jaccard index) | 0.21 |
| Collision rate — lane 3 / lane 4 / lane 5 | 19.2% / 23.1% / 18.9% |
| Closing-acceleration-driven collision rate — lane 5 / lane 4 / lane 3 | 5.1% / 3.7% / 3.4% |
| Discriminator range across the 27-setting parameter grid | 0.3%–66.4% (median ≈ 15%) |
| Discriminator marginal mean by leader braking rate (4 / 6 / 8 m/s$^2$) | 10.3% / 22.4% / 30.3% |
| Discriminator marginal mean by reaction time (0.5 / 1.0 / 1.5 s) | 6.1% / 18.2% / 38.7% |
| Discriminator marginal mean by follower braking rate (4 / 6 / 8 m/s$^2$) | 34.5% / 17.7% / 10.9% |
| Spearman correlation vs. MTTC, range across parameter grid | 0.36–0.62 |
| Lane risk-ranking correlation vs. central setting, range across grid | 0.07–1.00 (median 0.42) |
| Closing-acceleration-driven lane-ranking correlation vs. central setting, median | 0.71 |
| Minimum-discriminator parameter setting | $b_L$ = 4 m/s², $t_r$ = 0.5 s, $b_F$ = 8 m/s² (0.32%) |
| Maximum-discriminator parameter setting | $b_L$= 8 m/s², $t_r$ = 1.5 s, $b_F$ = 4 m/s² (66.33%) |

Algorithm A1. CCAR computation pipeline.

| |
|---|
| **INPUT**: trajectory table with columns |
| [vehicleID, frame, time, lane, position, speed, acceleration, leaderID, leaderGapDistance] |
| parameters: b_L, t_r, b_F (assumed leader brake, reaction time, follower brake capability) |
| |
| **STEP 1** — Data quality audit |
| for each vehicle, sorted by frame: |
| v_hat <- central_difference(position, time) |
| a_hat <- central_difference(speed, time) |
| compare v_hat to stored speed, a_hat to stored acceleration via RMSE |
| flag frames with \|acceleration\| beyond physical bound |
| |
| **STEP 2** — Leader–follower pairing |
| for each follower frame: |
| join the leader's speed and acceleration at the same frame using leaderID |
| exclude frames with: no leader, non-positive gap, implausible acceleration |
| |
| **STEP 3** — Baseline measures (per frame) |
| TTC <- gap / (followerSpeed - leaderSpeed) if closing, else infinity |
| DRAC <- (followerSpeed - leaderSpeed)^2 / (2*gap) if closing, else 0 |
| MTTC <- smallest positive root of the relative-motion quadratic incorporating both vehicles' accelerations |
| |
| **STEP 4** — CCAR kernel (per frame) |
| alpha <- max(followerAcceleration, 0) |
| T_L <- leaderSpeed / b_L // leader stops here |
| w_1 <- followerSpeed + alpha * t_r // speed entering braking |
| d_1 <- followerSpeed*t_r + 0.5*alpha*t_r^2 // distance during reaction |
| define leader_displacement(tau): |
| if tau <= T_L: leaderSpeed*tau - 0.5*b_L*tau^2 |
| if tau <= T_L: leaderSpeed*tau - 0.5*b_L*tau^2 |
| define follower_displacement(tau): |
| if tau <= t_r: followerSpeed*tau + 0.5*alpha*tau^2 |
| else: d_1 + w_1*(tau-t_r) - 0.5*b_F*(tau-t_r)^2 (capped once follower speed reaches zero) |
| define projected_gap(tau): |
| gap_0 + leader_displacement(tau) - follower_displacement(tau) |
| evaluate projected_gap at the candidate times: |
| { 0, phase-1 stationary point, t_r, T_L, phase-2 stationary point, follower-stop time } |
| s_min <- minimum of projected_gap over these candidates |
| R <- max(0, 1 - s_min / gap_0) |
| collision <- (s_min <= 0) |
| if collision: severity <- closing speed at first tau where gap = 0 |
| |
| **STEP 5** — Discriminator (RQ1) |
| discriminator_set <- frames where (TTC >= safe_threshold) AND collision |
| |
| **STEP 6** — Incremental information (RQ2) |

| compute Spearman rank correlation of R against TTC, DRAC, MTTC |
|---|
| compute Jaccard overlap of top-decile-risk frames: CCAR vs MTTC partition frames into narrow gap x speed cells; |
| within each cell, test whether R rises with alpha |
| |
| **STEP 7** — Spatial characterisation |
| group frames by lane; report mean R, collision rate, |
| and alpha-driven-collision rate per lane |
| (alpha-driven = collision under measured alpha but not under alpha=0) |
| |
| **STEP 8** — Robustness sweep (RQ4) |
| for each combination of b_L, t_r, b_F in their swept ranges: repeat Steps 4-7 |
| report the range and central tendency of each headline statistic across all combinations |
| |
| **OUTPUT**: per-frame CCAR risk table; discriminator statistics; |
| correlation and overlap statistics; spatial summary; |
| robustness summary across the parameter grid |